\documentclass[reprint,twocolumn,aps,prl,superscriptaddress,longbibliography]{revtex4-2}
\usepackage{mhchem}
\usepackage{float}
\usepackage[english]{babel}
\usepackage[utf8]{inputenc}
\usepackage{fancyhdr}
\usepackage{sidecap}
\usepackage{multirow}

\usepackage{graphicx}
\usepackage{tabularx}
\usepackage{braket}
\usepackage{tikz}
\usepackage{color}
\usepackage{amsmath}
\usepackage[colorlinks=True, linkcolor=blue, filecolor=magenta, urlcolor=blue,citecolor=blue]{hyperref}
\begin{document}
%%%%%%%%% Title %%%%%%%%%%%%%%%%%%%%%%%%%%%%%%%%%%%%%%%%%%%%%%%%%%%%%%%%%%%%%%
\title{\Large{\textbf {Growth of ultra-clean single crystals of RuO$_2$}}}
%%%%%%%%% Authors name%%%%%%%%%%%%%%%%%%%%%%%%%%%%%%%%%%%%%%%%%%%%%%%%%%%%%%%%	
\author{Shubhankar Paul}
\email{paul.shubhankar.52x@st.kyoto-u.ac.jp, shubhp@iitk.ac.in}
\affiliation{Toyota Riken–Kyoto University Research Center (TRiKUC), Kyoto 606-8501, Japan}
\affiliation{Department of Electronic Science and Engineering, Graduate School of Engineering, Kyoto University, Kyoto 615-8510, Japan}
\affiliation{Department of Physics, Indian Institute of Technology Kanpur, Kanpur 208016, India}

\author{Giordano Mattoni}
\affiliation{Toyota Riken–Kyoto University Research Center (TRiKUC), Kyoto 606-8501, Japan}

\author{Hisakazu Matsuki}
\affiliation{Toyota Riken–Kyoto University Research Center (TRiKUC), Kyoto 606-8501, Japan}
\author{Thomas Johnson}
\affiliation{Toyota Riken–Kyoto University Research Center (TRiKUC), Kyoto 606-8501, Japan}
\author{Chanchal Sow}
\affiliation{Department of Physics, Indian Institute of Technology Kanpur, Kanpur 208016, India}

\author{Shingo Yonezawa}
\affiliation{Department of Electronic Science and Engineering, Graduate School of Engineering, Kyoto University, Kyoto 615-8510, Japan}

\author{Yoshiteru Maeno}
\email{maeno.yoshiteru.b04@kyoto-u.jp}
\affiliation{Toyota Riken–Kyoto University Research Center (TRiKUC), Kyoto 606-8501, Japan}

\date{\today}
%%%%%%%%%%%%  abstract  %%%%%%%%%%%%%%%%%%%%%%%%%%%%%%%%%%%%%%%%%%%%%%%
\begin{abstract}
We report the details of the growth of ultra-clean single crystals for RuO$_2$, a candidate material for altermagnetism. 
By using a crystal-growth tube with a necking structure and precisely controlling the conditions of the sublimation transport method, it is possible to control the morphology of the crystals.
We obtained crystals in mainly three kinds of morphologies: thick plate-like crystals typically $5\times 3\times 2 \mathrm{mm}^3$ and up to $10\times 5\times 2 \mathrm{mm}^3$ with a large (101) facet, rhombohedral columnar crystals elongating along the [001] direction, and fiber and needle crystals of length up to 8 mm and the width of 0.1 - 0.4 mm.
These crystals show residual resistivity of about $30\ \mathrm{n} \Omega \mathrm{cm}$ and a residual resistivity ratio (RRR) up to 1200.
The crystals do not exhibit any signs of magnetic ordering down to low temperatures.
\end{abstract}
\maketitle

%%%%%%%%%%% Introduction  %%%%%%%%%%%%%%%%%%%%%%%%%%%%%%%%%%%%%%%%%%%%%%	
\section*{Introduction}
The availability of high-quality single crystals is essential for the correct interpretation of novel phenomena in quantum materials.
RuO$_2$ in the rutile structure has long been known as a good metal \cite{guthrie1931magnetic} and has been used in a wide variety of applications, such as catalyst for the oxygen evolution reaction (OER)~\cite{Over2000catalyst, Ping2024catalyst} and low-temperature thermometers with low magnetoresistance \cite{Doi_RuO2-sensor_LT17-1984}.
Under biaxial strain applied by a substrate, superconductivity emerges in the films of RuO$_2$ \cite{ruf2021strain, uchida2020superconductivity}.

The intrinsic magnetic ground state of RuO$_2$ is under current debate. It was known as a paramagnetic metal until exotic antiferromagnetic (AFM) ordering with N\'eel temperature $T_\textrm{N}>$ 300 K was reported by neutron scattering \cite{berlijn2017itinerant} and subsequently confirmed by resonant X-ray scattering \cite{zhu2019anomalous} as well as by angle-resolved photoemission spectroscopy (ARPES)~\cite{Lin2024ARPES}.  
The AFM state of RuO$_2$ attracts much research attention because of a direction-dependent exchange energy splitting which is the largest among the recently proposed altermagnetic candidate materials \cite{vsmejkal2022beyond,vsmejkal2022emerging}.

Altermagnets were proposed as a class of AFM materials that can be considered to be the third class of magnetic materials distinct from ferromagnetic (FM) and conventional AFM materials~\cite{vsmejkal2022beyond, vsmejkal2022emerging}.
They are expected to exhibit novel properties such as anisotropic spin splitting~\cite{ahn2019antiferromagnetism}, anomalous Hall effect~\cite{vsmejkal2020crystal}, spin current~\cite{naka2019spin}, and characteristic magnon excitations~\cite{naka2019spin}.
Contrary to reports of an AFM ordered state of RuO$_2$, more recent reports demonstrate a paramagnetic ground state \cite{Kiefer2025,Kessler2024RuO2, osumi2025ARPES,wu-eaton2025q-osc}.
On the other hand, anomalous Hall effect in RuO$_2$ films supports the AFM state~\cite{Tschirner2023APLMatt}.
Therefore, reliable data using high-quality single crystals is required to conclude on the intrinsic magnetic ground state of RuO$_2$.

There have been a number of reports on the growth of bulk single crystals of RuO$_2$~\cite{Schfer1963, Butler1971, huang1982growth, mertig1986specific, lin2004low, Chen2004raman, pawula2024multiband,  Kiefer2025}.
Most of these reports use the chemical tranport reaction method, specifically the sublimation-crystallization method, as used in this study.
RuO$_2$ and/or Ru are used as starting materials and the volatile RuO$_3$ gas is transported, decomposed, and crystalized into single crystals of RuO$_2$~\cite{Lin2024ARPES}.  
The reported size of the crystal varies from sub-mm to about 10 mm.
Nanowire crystals of 100-nm thickness and 10$\mu$m length along the [001] direction were also reported \cite{Kim2010nanowire}. 
A common index used to characterize the quality of the crystals, the residual resistivity ratio (RRR), ranges from 12$\sim$100 \cite{lin2004low} to  
about 500 \cite{mertig1986specific}.

In this report, we describe the details of the growth and basic characterization of ultra-clean single crystals of RuO$_2$ with the RRR exceeding 1000.
Depending systematically on the growth conditions, three kinds of crystals are obtained: namely flat shape with the (101) facet, rhombohedral columnar shape elongating along the [001] direction, and fiber/needle shape along the [001] direction with a length up to 8 mm.

%%%%%%%%%%%%%%%%%%%%%%%%%%%%%%%%%%%%%%%%%%%%%%%%%%
\section*{Single Crystal Growth}
\subsection*{Preparation for the growth}

We use the sublimation transport method~\cite{huang1982growth} with RuO$_2$ as a starting material. 
The starting powder of RuO$_2$ (Rare Metallic Ltd., 3N containing less than 3 ppm sodium) is compacted into a 6-mm-diameter cylinder.
According to the Glow Discharge Mass Spectroscopy (GDMS) analysis performed over 72 elements, the powder typically contains $\sim$ 300 ppm Cl, $\sim$ 50 ppm Si, and $\sim$ 20 ppm each for B, Fe, and Cu as main impurities.  
Thus, we may consider it as $99.95\%$ pure.

%%%%%%%% Fig. 1 %%%%%%%%%%%%%%%%%%%%%%%%%%%%%%%
 \begin{figure}[t]
   \begin{center}
   \includegraphics[scale=0.28]{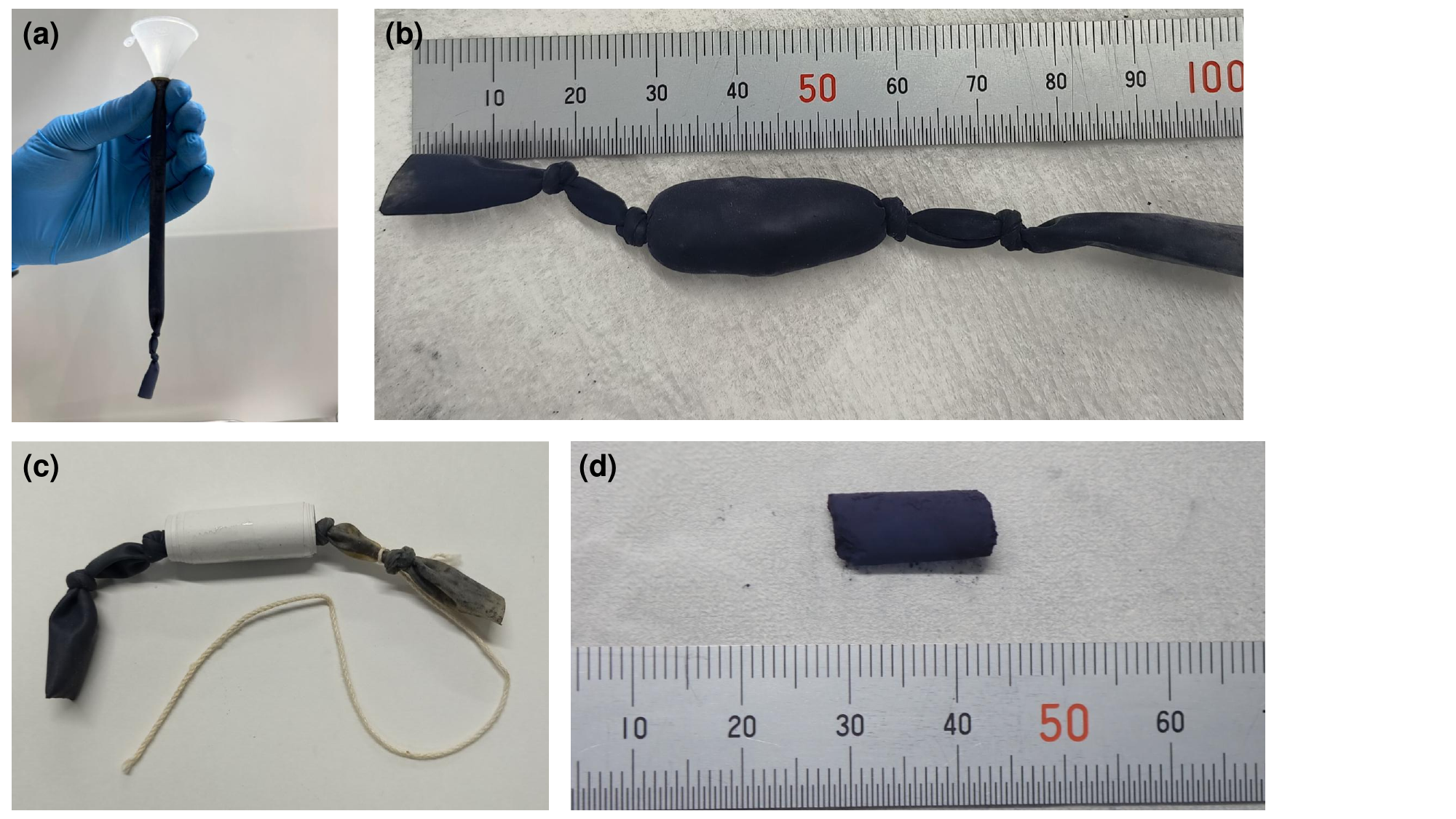}
   \caption{Preparation of the starting material cylinder of RuO$_2$. (a) Introducing RuO$_2$ powder into latex balloon. (b) Latex balloon filled with 6.7 g of RuO$_2$ powder. (c) RuO$_2$ powder in the latex  balloon is inserted in a paper tube of 6-mm diameter. It is ready to be compressed in a hydraulic press. (d) Compacted RuO$_2$ cylinder of 6.3 g after taken out of the latex balloon.}
   \label {fig1}
   \end{center}
 \end{figure}
%%%%%%%%%%%%%%%%%%%%%%%%%%%%%%%%%%%%%%%%%%%%%%
We avoid using a metallic pelletizer to minimize contamination and use a latex sleeve to prepare the RuO$_2$ cylinder, in a similar way used for the growth of Sr$_2$RuO$_4$~\cite{Bobowski2019}.
A latex sleeve (eastsidemed \textregistered, IMG $\phi$\ 7 mm) is cut in half to the length of 16 cm.
Both inner and outer surfaces are wiped with ethanol and ultrasonicated in ethanol to remove the TiO$_2$ coating powder.
To flip the latex sleeve inside out, we use a 3-mm diameter glass rod and adhesive tape that is removed before sonication.
We coat both surfaces with RuO$_2$ powder to avoid sticking.
Then, as shown in Fig. \ref{fig1} (a) we fasten one side of the sleeve by tying two knots to form a balloon and introduce RuO$_2$ powder of 4 to 10 g with a plastic funnel.
After manually squeezing out excess air, we seal the other end of the balloon by tying two knots as in Fig. \ref{fig1} (b) and tie a cotton string.
We then introduce this powder-packed balloon in a paper tube into a 6-mm diameter paper tube (Fig. \ref{fig1} (c)) and compress it in a water-filled steel cylinder tube with a hydraulic press at 110 MPa for 15 min. 
To avoid contamination from iron in the hydraulic cylinder, we placed the balloon in a plastic tube with a bottom cap to avoid direct contact with the inner wall of the steel cylinder.
We cut the latex balloon with ceramic scissors to take out the compacted RuO$_2$ rod from the balloon (Fig. \ref{fig1} (d)) and place it at the inlet side of the crystal-growth tube of the furnace.
%%%%%%%%%%%%%%%%%%%%%%%%%%%%%
\subsection*{Furnace}
%%%%%%%%% Fig. 2 %%%%%%%%%%%
\begin{figure*}[t]
	\begin{center}
			\includegraphics[scale=0.5]{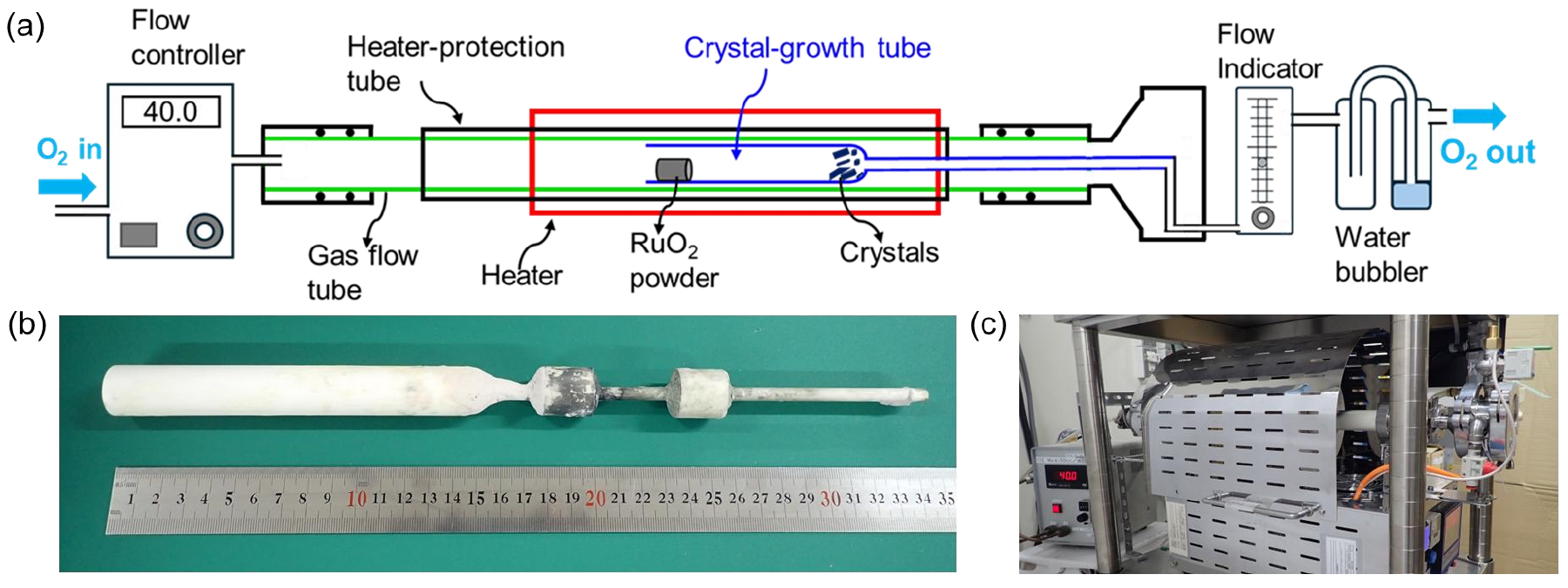}
		\caption{Crystal growth apparatus. (a) Schematic of the heater element (red), heater-protection tube (black), gas-flow tube (green), and crystal-growth tube (blue). A flow controller is placed at the inlet side. (b) Photo of the crystal-growth tube with two radiation-shield baffles. Only this tube is taken out of the furnace for each growth run. (c) Photo of the furnace with the flow controller.}
		\label {fig2}
	\end{center}
 \end{figure*}
 %%%%%%%%%%%%%%%%%%%%%%%%%%%
RuO$_2$ crystals were grown using the apparatus depicted in Fig. \ref{fig2} (a). 
Sublimation and crystallization takes place in the crystal-growth tube unit consisting of an alumina tube of 22-mm outer diameter (OD)/18-mm inner diameter (ID) with length 150 mm connected to a narrow alumina tube of 6-mm OD/3-mm ID, 180 mm in length. 
This unit is assembled using ceramic adhesive of high alumina content (Isolite Insulating Products, Kaostick) that withstands temperatures up to 1700$^{\circ}$C as shown in Fig. \ref{fig2} (b).
The unit is then inserted into a gas-flow tube of alumina of 30-mm OD/24-mm ID, placed in a single-zone tube furnace (Ful-Tech, FT-01VAC-1650).
This crystal-growth tube is readily removable for loading the starting RuO$_2$ rod and collecting the grown crystals from the furnace for each growth run.
All flowing oxygen gas (99.99\% purity), controlled at 40 cm$^3$/min by a gas-flow controller at the inlet of the gas-flow tube tube, goes through the inner crystal-growth tube.
The furnace with the gas-flow tube and the flow controller is shown in the photo of Fig. \ref{fig2} (c).

\subsection*{Growth conditions}
The growth conditions are summarized in Table \ref{tab1}. 
After placing the compacted rod of RuO$_2$ near the entrance of the crystal-growth tube in the middle of the furnace, the temperature of the furnace is raised to a target sublimation temperature in 1 hr and kept at that temperature typically for 65 to 100 hrs, then cooled to ambient temperature in 1 hr.
The temperature distribution is measured by sliding a R-type thermocouple into the crystal-growth tube from the narrow-tube side at the furnace temperature of 1100$^\circ$C, which is controlled by another R-type thermocouple placed outside of the heater element. 
It is confirmed that the temperature in the input area of the crystal growth tube in Fig. \ref{fig2} (a) matches well to the furnace temperature, but the temperature in the neck position is at 850$^\circ$C.   
Since the metallic sheath of our monitoring thermocouple withstands only up to 1200$^\circ$C, we denote in Table \ref{tab1} the furnace temperature as the sublimation temperature, and the growth temperature as more than 250$^\circ$C lower if the sublimation temperature is greater than 1100$^\circ$C. 
After each growth, the RuO$_2$ crystals, mostly found near the end of the 18-mm-ID tube, are collected by removing the crystal-growth tube from the furnace. 
Most of the crystals are readily removed by mechanically tapping over the crystal-growth tube.

%%%%%%%%%%%% Table %%%%%%%%%%%%
 \begin{table*}
    \small
    \caption{Conditions of the growth of RuO$_2$ crystals. Sublimation temperature is monitored by the system thermocouple; 
    growth temperature is defined in the text. The flow rate is the value as controlled at room temperature.}
    \label{tab1}
    \begin{tabularx}{1\textwidth}{clccccX}
 	\hline
        \hline
     Run  & Material  & Sublimation  & Growth  & Growth & Flow rate &  \\ [-1pt]
     number& (amount g)& temp. ($^{\circ}$C) & temp ($^{\circ}$C) & time (hrs) & (cm$^3$/min)  &  Crystal morphology  \\[+2pt]
        \hline 
     1 & RuO$_2$ \ (5.5) & 1100 & \;\;\;\;\;850 & \: 65  & 40 & Tiny crystals, 90$\%$ of the starting material remains.\\ [+2pt]
     2 & RuO$_2$ (10.3) & 1200 & \ $< 950$ & 168  & 40 & flat and columnar crystals  \\ [+2pt]
     3 & RuO$_2$ \ (8.7) & 1250 & $< 1000$ & 100  & 40 & flat and columnar crystals  \\ [+2pt]
     4 & RuO$_2$ \ (4.9) & 1300 & $< 1050$ & \: 65  & 40 & large flat, columnar, and fiber/long needle crystals\\  [+2pt]
     5 & RuO$_2$ \ (6.2) & 1300 & $< 1050$ & 100  & 40 & large flat, columnar, and fiber/long needle crystals \\ [+2pt]
     6 & RuO$_2$ \ (4.3) & 1325 & $< 1075$ & \: 65  & 40 & flat, columnar, and fiber/needle crystals  \\ [+2pt]
     7 & RuO$_2$ \ (5.6) & 1350 & $< 1100$ & 100  & 40 & flat, columnar, longer fiber/needle-twin crystals \\ [+2pt]
        \hline
        \hline
    \end{tabularx}
\end{table*}
%%%%%%%%%%%%%%%%%%%%%%%%%%%%%%%%%%%%%%%%%%%%%
\section*{Characterization of the Crystals}
\subsection*{Morphology of the crystals}
\begin{figure*}[t]
	\begin{center}
		\includegraphics[scale=0.6]{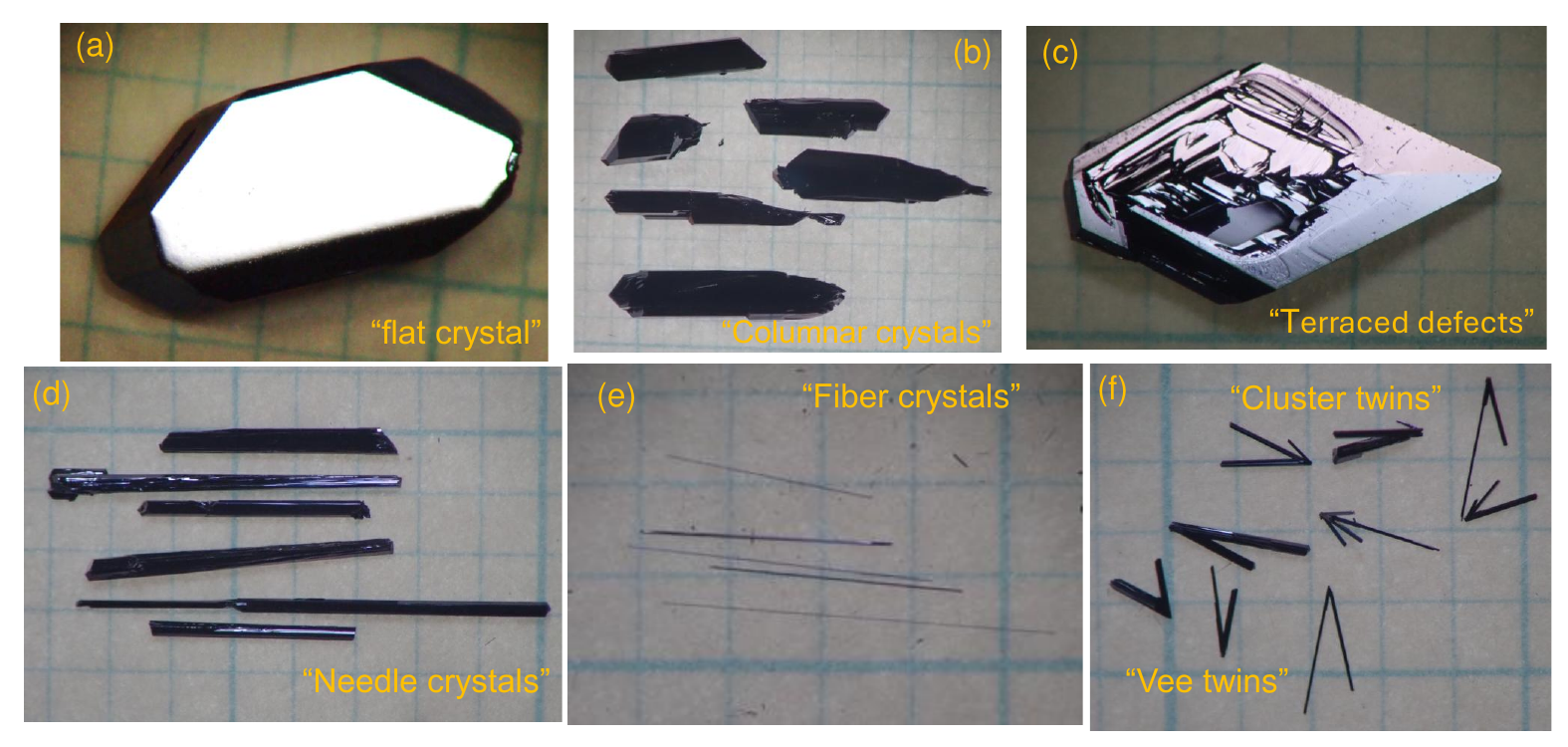}
		\caption{Optical images of the ``zoo'' of RuO$_2$ crystals in various morphologies: 
        (a) flat shape with a large (101) facet (1300$^\circ$C), 
        (b) rhombohedral columnar shape with the column axis along [001] (1300$^\circ$C), 
        (c) (101) facet showing terraced depression (1300$^\circ$C), 
        (d) needle shape up to 8 mm along [001] (1300$^\circ$C),
        (e) fiber along [001] with a square or triangular cross section (1350$^\circ$C),  
        (f) vee-/cluster-twin fiber crystals along [001] (1350$^\circ$C). 
        The background scale is 1 mm.
        The numbers in the parentheses indicate the sublimation temperature.}
        \label {fig3}
	\end{center}
 \end{figure*}

We observed a systematic variation in crystal morphology and size by changing the sublimation temperature and the associated growth temperature. 
Fig. \ref{fig3} represents these RuO$_2$ crystals with a variety of morphologies.
The crystals grow mainly in three kinds of shapes depending on the set temperature.
A sublimation temperature higher than 1200$^{\circ}$C is needed to promote the sublimation of the starting RuO$_2$ material.
With a sublimation temperature at 1250$^{\circ}$C, the crystals are mainly flat shape with a large (101) facet of up to 5$\times$3$\times$2 mm$^3$ and rhombohedral columnar (with the (110) and (100) side surfaces). 

At 1300$^{\circ}$C, we obtained larger crystals suitable for measurements of bulk physical properties.
Flat-shaped crystals grow in size up to $10\times 5\times 2 \mathrm{mm}^3$ with a large (101) facet as in Fig. \ref{fig3}(a).
Rhombohedral columnar crystals elongating along the [001] direction grow in size up to $8\times 2\times$ 1 $\mathrm{mm}^3$ as represented in Fig. \ref{fig3}(b).
In flat and columnar crystals, we often observe craters on the facet surface consisting of step-like terraced defects represented in Fig. \ref{fig3} (c).
In addition, many needle-shaped crystals up to 8 mm along the $c$-axis are grown with the side surfaces of (100) planes of width 0.2 $\sim$ 0.4 mm (Fig. \ref{fig3} (d)). 
There are also fine fiber crystals, which become longer at 1350$^{\circ}$.

At 1350$^{\circ}$C, we obtained fine fiber crystals longer than 4 mm, with small cross section much less than 100$\times$100 $\mu\mathrm{m}^2$.
They have either square or bilateral-triangular cross-sections; the square side consists of (100) facets and the triangular side consists of one (100) and two (110) facet planes as shown in \ref{fig3}(e), as identified by Laue photos.
Raising the temperature to 1350°C dramatically accelerates sublimation: RuO$_2$ evaporates more vigorously from the source in the form of RuO$_3$~\cite{Butler1971, huang1982growth} and condenses even at the far end of the 3-mm ID tube. 
Since the mass flow rate is conserved, the speed of the gas is much faster in the high temperature region and especially when the gas enters the narrow tube at the necking point.

%%%%%%%%%%  Fig. 3 %%%%%%%%%%%%%%
%%%%%%%%%%%%%%%%%%%%%%%%%%%%%%
At this temperature, we also observed distinct twin morphologies in fiber and needle crystals, similar to what were previously reported from lower-temperature, long-time (1200$^{\circ}$C, 15-days) growth~\cite{Butler1971}:
One is Vee twins (Fig. \ref{fig3}(f)), characterized by two fiber segments diverging at an acute angle of 23.5$^{\circ}$ and 55.5$^{\circ}$. 

The other is cluster twins, in which multiple fibers grow from a common nucleation point with a typical angle of 23.5$^{\circ}$. 

%%%%%%%%%%%%%%%%%%%%%%%%%%%%%%
\subsection*{Laue and XRD pattens}
\begin{figure}[b]
    \begin{center}
    \includegraphics[scale=0.65]{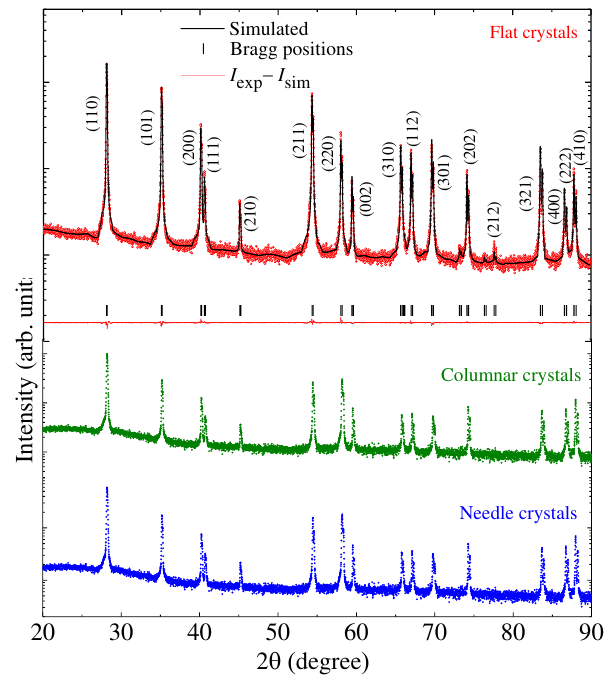}
    \caption{Powder XRD spectra of crushed crystals of RuO$_2$ in different morphologies but all grown with the sublimation temperature at 1300$^{\circ}$C. 
    Top: flat crystals (red), along with the simulated pattern (black). 
    The difference between the experimental and simulated intensities are also shown.
    Middle: columnar crystals; Bottom: needle crystals.}
    \label {fig4}
    \end{center}
 \end{figure}

We present in Fig. \ref{fig4} powder X-ray diffraction (XRD) spectra of crushed crystals in three morphologies.
%%%%%%%%%% Fig. 4 XRD  %%%%%%%%%%%%%
 %%%%%%%%%%%%%%%%%%%%%%%%%%%%%%%
We used a desk-top XRD spectrometer (RIGAKU, Miniflex 600-C) with Cu-K$\alpha$ radiation and the spectra measured down to $2\theta$ = 3$^\circ$ indicate no impurity phase.
We performed Rietveld structural profile refinement using the FullProf suite \cite{rodriguez1993recent}.
The results confirm the phase purity for the rutile RuO$_2$ with the lattice parameters $a = 4.49$  and $c = 3.11\  \mathrm{\AA}$, consistent among the crystals in different morphologies and consistent with previous reports~\cite{Butler1971, huang1982growth, Kiefer2025}.

\begin{figure*}
	\begin{center}
	\includegraphics[scale=0.55]{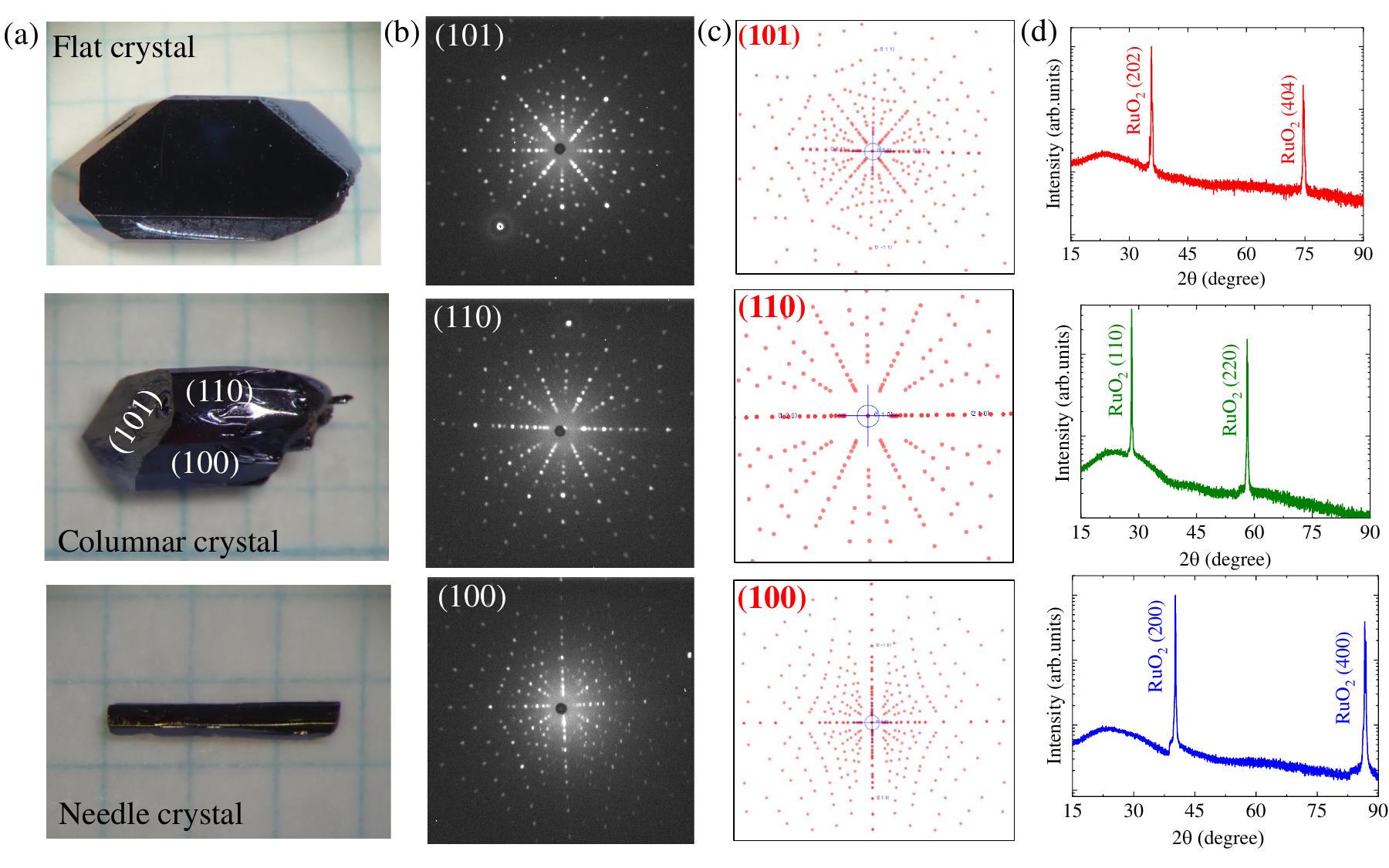}
	\caption{Comparison of RuO$_2$ crystals in three kinds of morphologies. 
        Top row: Flat crystal, middle row: Rhombohedral columnar crystal, bottom row: Needle-shape crystal. 
        (a) Optical images,
        (b) measured Laue photos of (101), (110), and (100) facets of the corresponding three crystals in (a), 
        (c) simulated Laue patterns,
        and (d) XRD spectra of the corresponding surfaces of the crystals.
        The flat crystal shown at the top of (a) is identical to the one in Fig. \ref{fig3}(a), but under different illumination.}
		\label {fig5}
	\end{center}
 \end{figure*}

Figure \ref{fig5}(a) represents the optical images of the flat, columnar, and needle-shaped crystals selected for characterization, and Fig. \ref{fig5}(b) is the back-scattered Laue diffraction patterns (RIGAKU, RASCO-BL$\rm I \hspace{-.01em}I $) of the facet surfaces of these three kinds of crystals.
The corresponding simulation patterns shown in Fig. \ref{fig5}(c) are generated using the Q-Laue software ($\copyright$ Stuart B. Wilkins, 2007); excellent agreement between experiments and simulations identifies that the large flat facet is the (101) plane, the columnar crystal has the (110) facet on the side surface, while the fiber crystal exhibits the (100) facet surface plane. 
XRD spectra taken from these RuO$_2$ crystals shown in Fig. \ref{fig5}(d) confirm these assignments.

 %%%%%%%%%%  Fig. 5  Laue %%%%%%%%%%%%%%
 
\begin{figure}
    \begin{center}
    \includegraphics[scale=0.3]{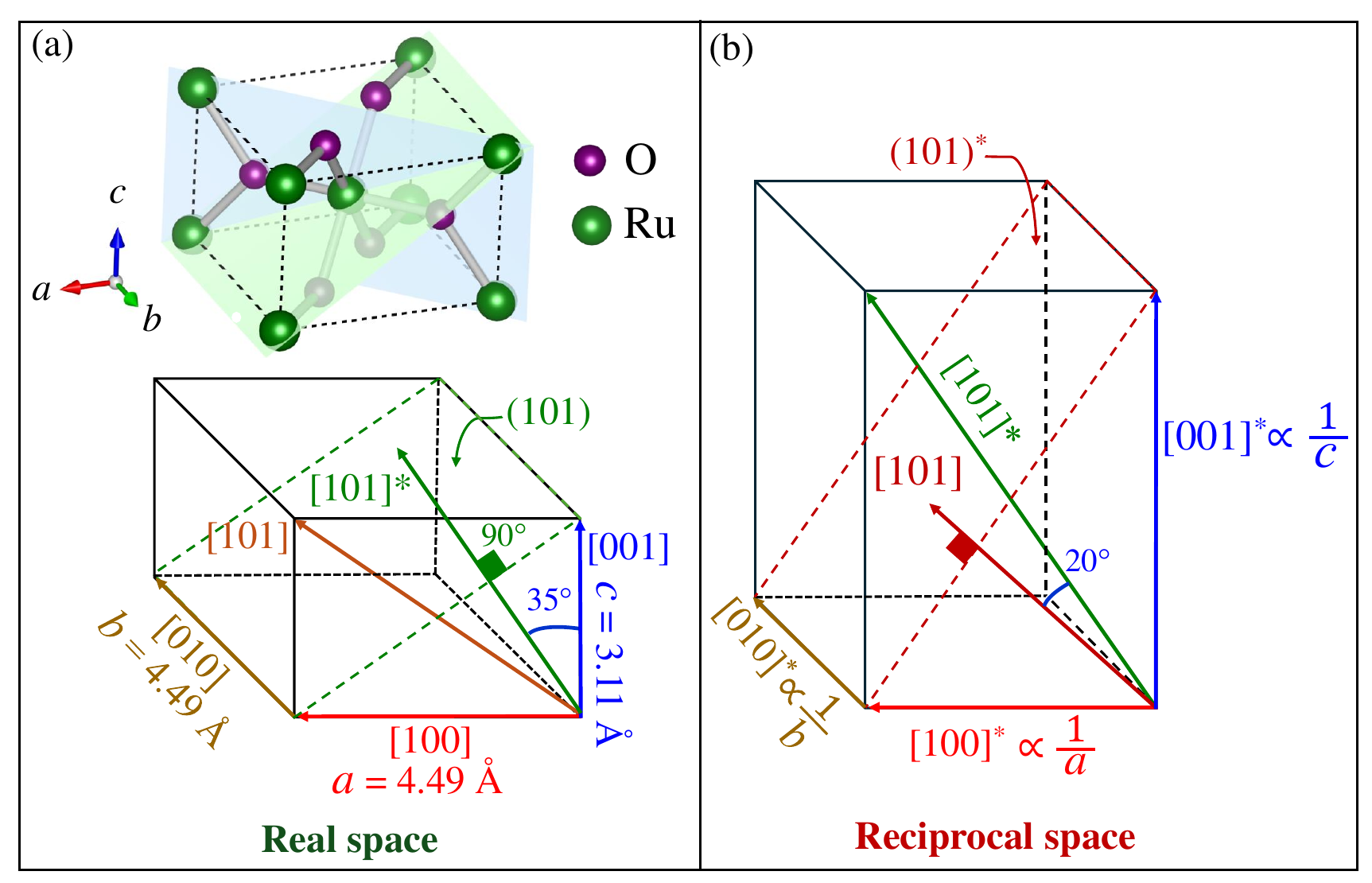}
    \caption{Schematic representation of the rutile RuO$_2$ structure with the (101) plane and the [101]* direction. (a) In real space with atomic arrangements~\cite{momma2008vesta} and (b) in reciprocal spaces.}        
    \label {fig6}
    \end{center}
 \end{figure}

Here we note a simple but confusing fact that the [101] direction is not perpendicular to the (101) plane in the tetragonal structure.
Figure \ref{fig6} shows the schematic representation of tetragonal rutile crystal structure of RuO$_2$ in real and reciprocal spaces. 
The asterisk (*) indicates the direction in reciprocal space.
The [101]* direction in real space is perpendicular to the (101) plane, which is tilted by 20$^\circ$ from the [101] direction. 

%%%%%%%%%% Fig. 6  %%%%%%%%%%%%%
 %%%%%%%%%%%%%%%%%%%%%%%%%%%%%%%%%%

%%%%%%%%%%%%%%%%%%%%%%%%%%%%%%%%%%%
\subsection*{Resistivity}
\begin{figure}[h]
   \begin{center}
   \includegraphics[scale=0.35]{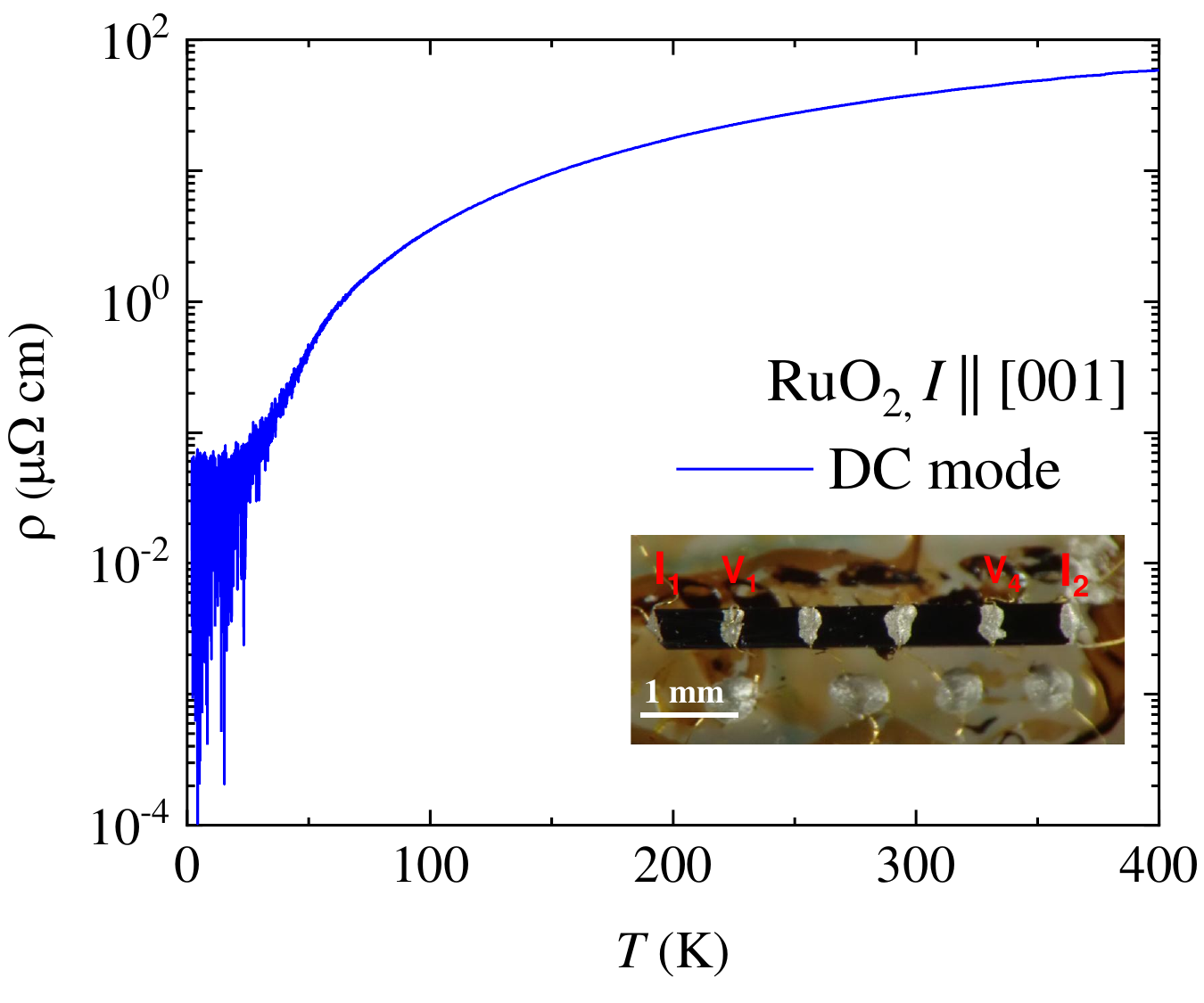}
   \caption{Resistivity of RuO$_2$ measured by the DC delta-mode method for the current along the [001] direction.
   The voltage terminals of V$_1$V$_4$ in Fig. \ref{fig7}(a) are used. See text for the details of the measurement method.}
   \label {fig7}
   \end{center}
\end{figure}

To evaluate the purity of the crystals, we measured the resistivity using both direct current (DC) and alternating current (AC) four-probe methods. 
Figure \ref{fig7} shows the resistivity between 1.8 and 400 K for the DC four-probe method, in which we used a delta-mode square-wave current of 10 mA with a cycle period of 50 ms.
A current source (Keithley, model 6221) and a nanovoltmeter (Keithley. model 2182) were configured in this delta mode.
The DC method usually yields accurate resistance values but due to a high conductivity of RuO$_2$, the measurements at low temperatures suffer from the noise floor of our instrument. 

\begin{figure}[ht]
   \begin{center}
   \includegraphics[scale=0.65]{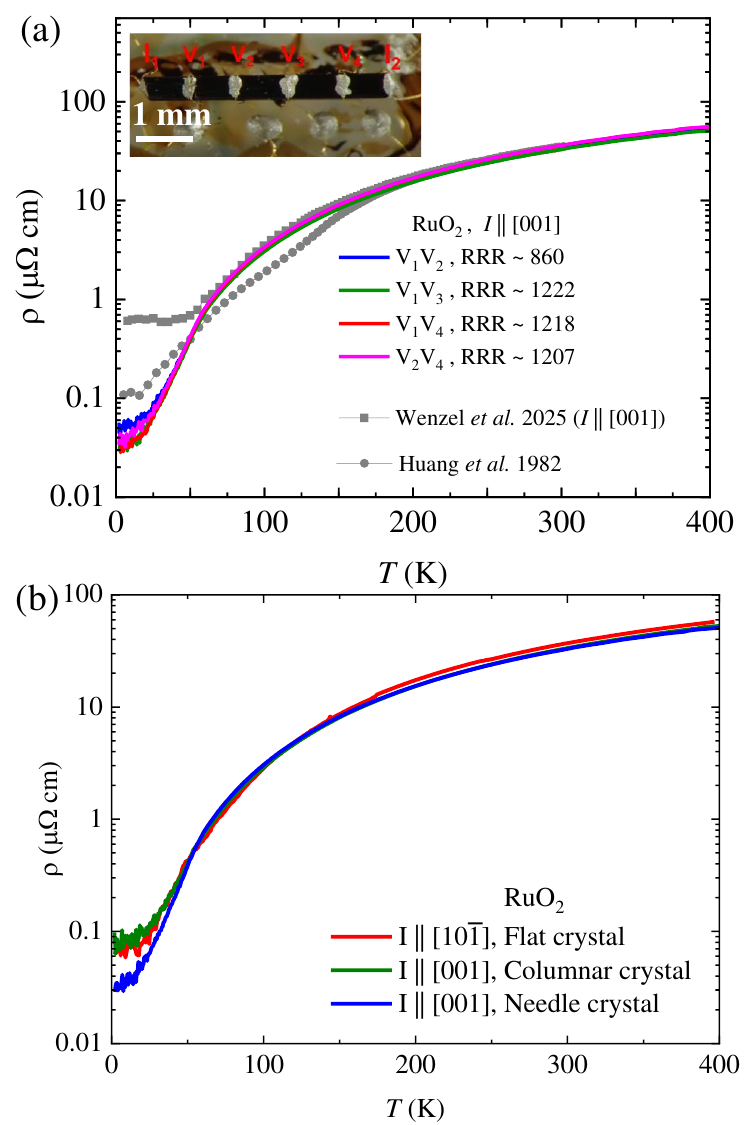}
   \caption{Resistivity of RuO$_2$ measured by the AC four-probe method. 
   (a) For the current along the [001] direction measured with different sets of the voltage leads to confirm the accuracy of the resistivity values. 
   Previous results reporting consistent resistivity values at room temperature are plotted for comparison. 
   Inset: Optical image of the fiber crystal with the current leads I$_1$ and I$_2$, and four voltage leads V$_1$ - V$_4$. 
   (b) Resistivities of crystals in different morphologies are consistent with each other, although there are some variations in the RRR between 560 and 1200.}
   \label {fig8}
   \end{center}
\end{figure}

Thus, we mainly used the AC four-probe method with a lock-in amplifier (Stanford Research, SR830) with its internal oscillator as a voltage source to drive a home-made voltage-to-current converter.
The amplitude and phase of the input current was monitored by a series standard resistor of 1.4 $ \Omega$.
We used a low frequency of 17 Hz to suppress the phase shift between the standard resistor and the sample at low temperatures to a negligible level.
As shown in the inset of Fig. \ref{fig8}, for current along the [001] direction, we attached a current lead at each end and four voltage leads on the side surface of a needle-like crystal using silver epoxy (EPO-TEK, H20E) cured at a high temperature of 500$^{\circ}$C to evaporate the epoxy content.
In this way, we obtained a mechanically robust contact with a much reduced contact resistance as low as 1 $\Omega$.
With a root-mean-square (RMS) current of 10 mA, the heating at each contact contact is on the order of $0.1\ \mathrm{mW}$ and negligible in this study.
The resistivity values from the AC method agree well with those by the DC method as demonstrated in Figs. \ref{fig7} and \ref{fig8}.

Figure \ref{fig8} (a) compares the resistivity along the $c$-axis between 1.8 and 400 K using different voltage-contact pairs.
To extract the resistivity accurately, we used a series of electrodes with different separation measured as the center-to-center distance of the silver contacts. 
This allowed us to obtain reproducible resistivity and RRR values, demonstrating the accuracy of the measurements.
As a comparison, we also include in Fig. \ref{fig8}(a) the resistivity reported in Refs. \cite{huang1982growth, wenzel2025fermi}, which were measured up to 300 K.
Their resistivity values at 300 K are consistent with our results of 36.2 $\mu\Omega$ cm by the AC method using V$_1$-V$_4$.
The obtained residual resistivity 0.03 $\mu \Omega$ cm at 1.8 K corresponds to a mean-free-path of 0.5 $\mu$m. 
This yields RRR $=$ 1200, higher than previous reports~\cite{huang1982growth, lin2004low, pawula2024multiband, wenzel2025fermi, Kiefer2025}, indicating high purity of the RuO$_2$ crystals in this report.
In accordance with the electronic structure that has a three-dimensional (3D) Fermi surface \cite{ahn2019antiferromagnetism, osumi2025ARPES, wu-eaton2025q-osc}, the resistivity is rather isotropic and does not vary significantly among crystals in different morphologies as shown in Fig. \ref{fig8}(b).

%%%%%%%%%%% Fig. 7 %%%%%%%%%%%%%
%%%%%%%%%%%%%%%%%%%%%%%%%%%%%%%

%%%%%%%%%%% Fig. 8 %%%%%%%%%%%%%
%%%%%%%%%%%%%%%%%%%%%%%%
%%%%%%%%%%%%% Fig. 9 %%%%%%%%%%%%%%%%
\begin{figure}[ht]
	\begin{center}
	\includegraphics[scale=0.35]{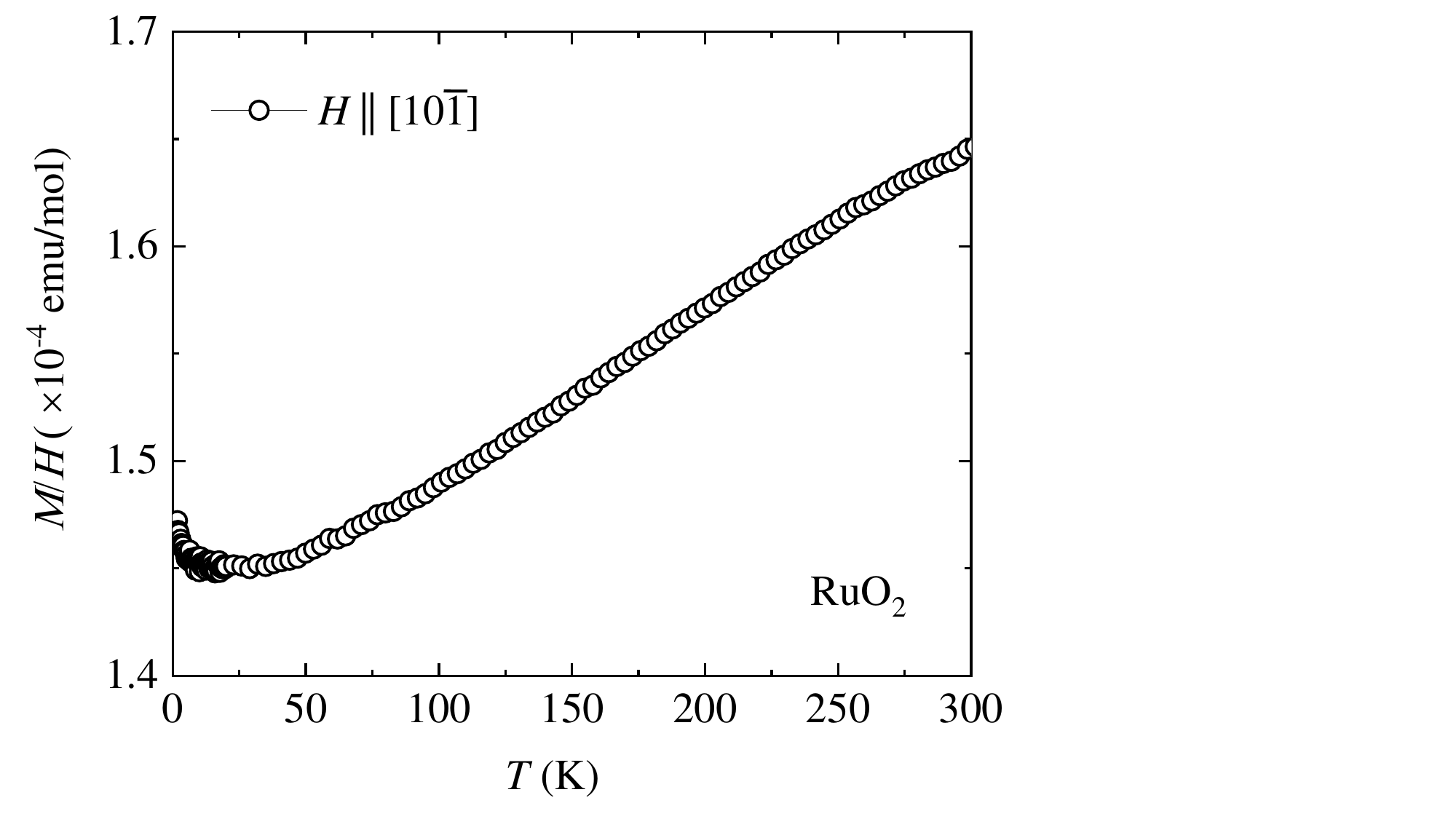}
	\caption{DC susceptibility $M/H$ of a RuO$_2$ crystal of the flat-facet shape down to 1.8 K under the field of 1 T along the [10$\bar{1}$] direction.
        Note that $1 \times 10^{-4}$ emu/mol for RuO$_2$ corresponds to 0.664 $\times 10^{-4}$ for dimensionless DC susceptibility in SI.}
	\label {fig9}
	\end{center}
\end{figure}
%%%%%%%%%%%%%%%%%%%%%%%%%%%%%%%%%%%%%%%
%%%%%%%%%%%%%%%%%%%%%%%%
\subsection*{Magnetization}
Fig. \ref{fig9} shows the DC susceptibility, magnetization $M$ divided by the magnetic field $H$, at 1 T from 1.8 to 300 K, measured using a superconducting quantum interference (SQUID) system (Quantum Design, MPMS-VL).

It is characterized by a Pauli paramagnetism with a positive temperature coefficient.
The low-temperature upturn in the present work is significantly smaller than previous reports~\cite{wenzel2025fermi, Kiefer2025}. 
This upturn is attributed to Curie paramagnetism arising from impurities with concentration in the range of just a few ppm.

\section*{DISCUSSION AND CONCLUSION}
We described the details of the growth and basic characterization of ultra-pure RuO$_2$ crystals with the RRR exceeding 1000 and a small low-temperature upturn in DC magnetic susceptibility corresponding to a few-ppm level paramagnetic impurities.
We attribute the enhanced crystal size, controllable morphologies, as well as ultra-high purity to the meticulous preparation of cylindrical pellets of the starting material and the use of a crystal-growth tube with a neck that defines the location of the crystal growth.
The morphology of the crystals varies from flat-facet to columnar to fiber/needle shapes; it can be controlled by tuning the growth temperature.
We found that the range of growth temperatures for long needle-shape crystals is narrow, compared to the ranges for columnar and flat-facet crystals. 
At higher sublimation temperature, needle and fiber crystals also form vee-twin and cluster-twin crystals.
The present furnace configuration with the crystal-growth tube may be widely applicable to the growth of other materials.

\subsection*{DATA AVAILABILITY}
The data that support the findings of this study are available from the corresponding
author upon reasonable request.

%%%%%%%%%%%%%
\subsection*{REFERENCES}
\bibliography{Paul-Maeno_RuO2_Crystal_growth}

\begin{thebibliography}{10}
\expandafter\ifx\csname url\endcsname\relax
  \def\url#1{\texttt{#1}}\fi
\expandafter\ifx\csname urlprefix\endcsname\relax\def\urlprefix{URL }\fi
\providecommand{\bibinfo}[2]{#2}
\providecommand{\eprint}[2][]{\url{#2}}

\bibitem{guthrie1931magnetic}
\bibinfo{author}{Guthrie, A.~N.} \& \bibinfo{author}{Bourland, L.}
\newblock \bibinfo{title}{Magnetic susceptibilities and ionic moments in the
  palladium and platinum groups}.
\newblock \emph{\bibinfo{journal}{Physical Review}}
  \textbf{\bibinfo{volume}{37}}, \bibinfo{pages}{303} (\bibinfo{year}{1931}).

\bibitem{Over2000catalyst}
\bibinfo{author}{Over, H.} \emph{et~al.}
\newblock \bibinfo{title}{Atomic-scale structure and catalytic reactivity of
  the \ce{RuO2}(110) surface}.
\newblock \emph{\bibinfo{journal}{Science}} \textbf{\bibinfo{volume}{287}},
  \bibinfo{pages}{1474–1476} (\bibinfo{year}{2000}).

\bibitem{Ping2024catalyst}
\bibinfo{author}{Ping, X.} \emph{et~al.}
\newblock \bibinfo{title}{Locking the lattice oxygen in \ce{RuO2} to stabilize
  highly active {R}u sites in acidic water oxidation}.
\newblock \emph{\bibinfo{journal}{Nature Communications}}
  \textbf{\bibinfo{volume}{15}} (\bibinfo{year}{2024}).

\bibitem{Doi_RuO2-sensor_LT17-1984}
\bibinfo{author}{Doi, H.}, \bibinfo{author}{Narahara, Y.},
  \bibinfo{author}{Oda, Y.} \& \bibinfo{author}{Nagane, H.}
\newblock \bibinfo{title}{\ce{RuO2} sensor}.
\newblock In \emph{\bibinfo{booktitle}{Proc. LT17, Kahrsruhe}},
  \bibinfo{pages}{405} (\bibinfo{publisher}{North-Holland, Amsterdam},
  \bibinfo{year}{1984}).

\bibitem{ruf2021strain}
\bibinfo{author}{Ruf, J.~P.} \emph{et~al.}
\newblock \bibinfo{title}{Strain-stabilized superconductivity}.
\newblock \emph{\bibinfo{journal}{Nature Communications}}
  \textbf{\bibinfo{volume}{12}}, \bibinfo{pages}{59} (\bibinfo{year}{2021}).

\bibitem{uchida2020superconductivity}
\bibinfo{author}{Uchida, M.}, \bibinfo{author}{Nomoto, T.},
  \bibinfo{author}{Musashi, M.}, \bibinfo{author}{Arita, R.} \&
  \bibinfo{author}{Kawasaki, M.}
\newblock \bibinfo{title}{Superconductivity in uniquely strained \ce{RuO2}
  films}.
\newblock \emph{\bibinfo{journal}{Physical Review Letters}}
  \textbf{\bibinfo{volume}{125}}, \bibinfo{pages}{147001}
  (\bibinfo{year}{2020}).

\bibitem{berlijn2017itinerant}
\bibinfo{author}{Berlijn, T.} \emph{et~al.}
\newblock \bibinfo{title}{Itinerant antiferromagnetism in \ce{RuO2}}.
\newblock \emph{\bibinfo{journal}{Physical Review Letters}}
  \textbf{\bibinfo{volume}{118}}, \bibinfo{pages}{077201}
  (\bibinfo{year}{2017}).

\bibitem{zhu2019anomalous}
\bibinfo{author}{Zhu, Z.} \emph{et~al.}
\newblock \bibinfo{title}{Anomalous antiferromagnetism in metallic \ce{RuO2}
  determined by resonant x-ray scattering}.
\newblock \emph{\bibinfo{journal}{Physical Review Letters}}
  \textbf{\bibinfo{volume}{122}}, \bibinfo{pages}{017202}
  (\bibinfo{year}{2019}).

\bibitem{Lin2024ARPES}
\bibinfo{author}{Lin, Z.} \emph{et~al.}
\newblock \bibinfo{title}{Observation of giant spin splitting and $d$-wave spin
  texture in room temperature altermagnet \ce{RuO2}}.
\newblock \emph{\bibinfo{journal}{preprint arXiv: 2402.04995}}
  (\bibinfo{year}{2024}).

\bibitem{vsmejkal2022beyond}
\bibinfo{author}{{\v{S}}mejkal, L.}, \bibinfo{author}{Sinova, J.} \&
  \bibinfo{author}{Jungwirth, T.}
\newblock \bibinfo{title}{Beyond conventional ferromagnetism and
  antiferromagnetism: A phase with nonrelativistic spin and crystal rotation
  symmetry}.
\newblock \emph{\bibinfo{journal}{Physical Review X}}
  \textbf{\bibinfo{volume}{12}}, \bibinfo{pages}{031042}
  (\bibinfo{year}{2022}).

\bibitem{vsmejkal2022emerging}
\bibinfo{author}{{\v{S}}mejkal, L.}, \bibinfo{author}{Sinova, J.} \&
  \bibinfo{author}{Jungwirth, T.}
\newblock \bibinfo{title}{Emerging research landscape of altermagnetism}.
\newblock \emph{\bibinfo{journal}{Physical Review X}}
  \textbf{\bibinfo{volume}{12}}, \bibinfo{pages}{040501}
  (\bibinfo{year}{2022}).

\bibitem{ahn2019antiferromagnetism}
\bibinfo{author}{Ahn, K.-H.}, \bibinfo{author}{Hariki, A.},
  \bibinfo{author}{Lee, K.-W.} \& \bibinfo{author}{Kune{\v{s}}, J.}
\newblock \bibinfo{title}{Antiferromagnetism in \ce{RuO2} as $d$-wave
  {P}omeranchuk instability}.
\newblock \emph{\bibinfo{journal}{Physical Review B}}
  \textbf{\bibinfo{volume}{99}}, \bibinfo{pages}{184432}
  (\bibinfo{year}{2019}).

\bibitem{vsmejkal2020crystal}
\bibinfo{author}{{\v{S}}mejkal, L.},
  \bibinfo{author}{Gonz{\'a}lez-Hern{\'a}ndez, R.}, \bibinfo{author}{Jungwirth,
  T.} \& \bibinfo{author}{Sinova, J.}
\newblock \bibinfo{title}{Crystal time-reversal symmetry breaking and
  spontaneous {H}all effect in collinear antiferromagnets}.
\newblock \emph{\bibinfo{journal}{Science Advances}}
  \textbf{\bibinfo{volume}{6}}, \bibinfo{pages}{eaaz8809}
  (\bibinfo{year}{2020}).

\bibitem{naka2019spin}
\bibinfo{author}{Naka, M.} \emph{et~al.}
\newblock \bibinfo{title}{Spin current generation in organic antiferromagnets}.
\newblock \emph{\bibinfo{journal}{Nature communications}}
  \textbf{\bibinfo{volume}{10}}, \bibinfo{pages}{4305} (\bibinfo{year}{2019}).

\bibitem{Kiefer2025}
\bibinfo{author}{Kiefer, L.} \emph{et~al.}
\newblock \bibinfo{title}{Crystal structure and absence of magnetic order in
  single-crystalline \ce{RuO2}}.
\newblock \emph{\bibinfo{journal}{Journal of Physics: Condensed Matter}}
  \textbf{\bibinfo{volume}{37}}, \bibinfo{pages}{135801}
  (\bibinfo{year}{2025}).

\bibitem{Kessler2024RuO2}
\bibinfo{author}{Keßler, P.} \emph{et~al.}
\newblock \bibinfo{title}{Absence of magnetic order in \ce{RuO2}: insights from
  $\mu$sr spectroscopy and neutron diffraction}.
\newblock \emph{\bibinfo{journal}{preprint arXiv:2405.10820}}
  (\bibinfo{year}{2024}).

\bibitem{osumi2025ARPES}
\bibinfo{author}{Osumi, T.} \emph{et~al.}
\newblock \bibinfo{title}{Spin-degenerate bulk bands and topological surface
  states of \ce{RuO2}}.
\newblock \emph{\bibinfo{journal}{preprint arXiv:2501.10649}}
  (\bibinfo{year}{2025}).

\bibitem{wu-eaton2025q-osc}
\bibinfo{author}{Wu, Z.} \emph{et~al.}
\newblock \bibinfo{title}{The {F}ermi surface of \ce{RuO2} measured by quantum
  oscillations}.
\newblock \emph{\bibinfo{journal}{preprint arXiv:2503.20621}}
  (\bibinfo{year}{2025}).

\bibitem{Tschirner2023APLMatt}
\bibinfo{author}{Tschirner, T.} \emph{et~al.}
\newblock \bibinfo{title}{Saturation of the anomalous {H}all effect at high
  magnetic fields in altermagnetic \ce{RuO2}}.
\newblock \emph{\bibinfo{journal}{APL Materials}} \textbf{\bibinfo{volume}{11}}
  (\bibinfo{year}{2023}).

\bibitem{Schfer1963}
\bibinfo{author}{Sch\"{a}fer, H.}, \bibinfo{author}{Schneidereit, G.} \&
  \bibinfo{author}{Gerhardt, W.}
\newblock \bibinfo{title}{Zur chemie der platinmetalle. \ce{RuO2} chemischer
  transport, eigenschaften, thermischer zerfall}.
\newblock \emph{\bibinfo{journal}{Zeitschrift f\"{u}r anorganische und
  allgemeine Chemie}} \textbf{\bibinfo{volume}{319}},
  \bibinfo{pages}{327–336} (\bibinfo{year}{1963}).

\bibitem{Butler1971}
\bibinfo{author}{Butler, S.} \& \bibinfo{author}{Gillson, J.}
\newblock \bibinfo{title}{Crystal growth, electrical resistivity and lattice
  parameters of \ce{RuO2} and \ce{IrO2}}.
\newblock \emph{\bibinfo{journal}{Materials Research Bulletin}}
  \textbf{\bibinfo{volume}{6}}, \bibinfo{pages}{81–89}
  (\bibinfo{year}{1971}).

\bibitem{huang1982growth}
\bibinfo{author}{Huang, Y.}, \bibinfo{author}{Park, H.} \&
  \bibinfo{author}{Pollak, F.~H.}
\newblock \bibinfo{title}{Growth and characterization of \ce{RuO2} single
  crystals}.
\newblock \emph{\bibinfo{journal}{Materials Research Bulletin}}
  \textbf{\bibinfo{volume}{17}}, \bibinfo{pages}{1305--1312}
  (\bibinfo{year}{1982}).

\bibitem{mertig1986specific}
\bibinfo{author}{Mertig, M.}, \bibinfo{author}{Pompe, G.} \&
  \bibinfo{author}{Hegenbarth, E.}
\newblock \bibinfo{title}{Specific heat of \ce{Nb}-doped \ce{RuO2} single
  crystals}.
\newblock \emph{\bibinfo{journal}{Physica Status Solidi (b)}}
  \textbf{\bibinfo{volume}{135}}, \bibinfo{pages}{335--342}
  (\bibinfo{year}{1986}).

\bibitem{lin2004low}
\bibinfo{author}{Lin, J.-J.} \emph{et~al.}
\newblock \bibinfo{title}{Low temperature electrical transport properties of
  \ce{RuO2} and \ce{IrO2} single crystals}.
\newblock \emph{\bibinfo{journal}{Journal of Physics: Condensed Matter}}
  \textbf{\bibinfo{volume}{16}}, \bibinfo{pages}{8035} (\bibinfo{year}{2004}).

\bibitem{Chen2004raman}
\bibinfo{author}{Chen, R.} \emph{et~al.}
\newblock \bibinfo{title}{A comparative study of microstructure of \ce{RuO2}
  nanorods via raman scattering and field emission scanning electron
  microscopy}.
\newblock \emph{\bibinfo{journal}{Solid State Communications}}
  \textbf{\bibinfo{volume}{131}}, \bibinfo{pages}{349–353}
  (\bibinfo{year}{2004}).

\bibitem{pawula2024multiband}
\bibinfo{author}{Pawula, F.} \emph{et~al.}
\newblock \bibinfo{title}{Multiband transport in \ce{RuO2}}.
\newblock \emph{\bibinfo{journal}{Physical Review B}}
  \textbf{\bibinfo{volume}{110}}, \bibinfo{pages}{064432}
  (\bibinfo{year}{2024}).

\bibitem{Kim2010nanowire}
\bibinfo{author}{Kim, M.~H.} \emph{et~al.}
\newblock \bibinfo{title}{Growth direction determination of a single \ce{RuO2}
  nanowire by polarized raman spectroscopy}.
\newblock \emph{\bibinfo{journal}{Applied Physics Letters}}
  \textbf{\bibinfo{volume}{96}} (\bibinfo{year}{2010}).

\bibitem{Bobowski2019}
\bibinfo{author}{Bobowski, J.~S.} \emph{et~al.}
\newblock \bibinfo{title}{Improved single-crystal growth of \ce{Sr2RuO4}}.
\newblock \emph{\bibinfo{journal}{Condensed Matter}}
  \textbf{\bibinfo{volume}{4}}, \bibinfo{pages}{6} (\bibinfo{year}{2019}).

\bibitem{rodriguez1993recent}
\bibinfo{author}{Rodr{\'\i}guez-Carvajal, J.}
\newblock \bibinfo{title}{Recent advances in magnetic structure determination
  by neutron powder diffraction}.
\newblock \emph{\bibinfo{journal}{Physica B: Condensed Matter}}
  \textbf{\bibinfo{volume}{192}}, \bibinfo{pages}{55--69}
  (\bibinfo{year}{1993}).

\bibitem{momma2008vesta}
\bibinfo{author}{Momma, K.} \& \bibinfo{author}{Izumi, F.}
\newblock \bibinfo{title}{Vesta: a three-dimensional visualization system for
  electronic and structural analysis}.
\newblock \emph{\bibinfo{journal}{Applied Crystallography}}
  \textbf{\bibinfo{volume}{41}}, \bibinfo{pages}{653--658}
  (\bibinfo{year}{2008}).

\bibitem{wenzel2025fermi}
\bibinfo{author}{Wenzel, M.} \emph{et~al.}
\newblock \bibinfo{title}{Fermi-liquid behavior of nonaltermagnetic \ce{RuO2}}.
\newblock \emph{\bibinfo{journal}{Physical Review B}}
  \textbf{\bibinfo{volume}{111}}, \bibinfo{pages}{L041115}
  (\bibinfo{year}{2025}).

\end{thebibliography}
%%%%%%%%%%%%

\subsection*{ACKNOWLEDGEMENTS}
This work was supported by the JSPS KAKENHI (JP22H01168, JP23K22439) and the JST Sakura Science Exchange Program. 
S.P. and C.S. acknowledge SRG SERB Grants (SRG2019-001104, CRG-2022-005726, EEQ-2022-000883), India, and Initiation Grant (IITK-2019-037), IIT Kanpur, for financial support.
G.M. acknowledges support from the Kyoto University Foundation, JSPS KAKENHO (JP25K17346), 
and Toyota Riken Scholar Program.
T.J. acknowledges support as JSPS International Research Fellow (PE24047).
\\

\subsection*{AUTHOR CONTRIBUTIONS}
SP and YM designed and constructed the furnace, SP mainly grew the crystals with YM and HM, SP mainly measured the resistivity and magnetization with GM. SP and YM drafted the article and all the authors contributed to revise it. \\

\subsection*{COMPETING INTERESTS}
The authors declare no competing interests.

\subsection*{ADDITIONAL INFORMATION}
Correspondence and requests for materials should be addressed to Y.M. (maeno.yoshiteru.b04@kyoto-u.jp) and S.P. (paul.shubhankar.52x@st.kyoto-u.ac.jp, shubhp@iitk.ac.in).

\bibliographystyle{naturemag}
\end{document}